**Ultrafast Nano-Oscillators Based on Interlayer-Bridged Carbon Nanoscrolls**

Zhao Zhang[1], Teng Li[1,2,*]

[1]*Department of Mechanical Engineering, University of Maryland, College Park, MD 20742,USA*

[2]*Maryland NanoCenter, University of Maryland, College Park, MD 20742,USA*

**Abstract**

We demonstrate a viable approach to fabricating ultrafast axial nano-oscillators based on carbon nanoscrolls (CNSs) using molecular dynamics simulations. Initiated by a single-walled carbon nanotube (CNT), a monolayer graphene can continuously scroll into a CNS with the CNT housed inside. The CNT inside the CNS can oscillate along axial direction at a natural frequency of 10s gigahertz (GHz). We demonstrate an effective strategy to reduce the dissipation of the CNS-based nano-oscillator by covalently bridging the carbon layers in the CNS. We further demonstrate that, such a CNS-based nano-oscillator can be excited and driven by an external AC electric field, and oscillate at more than 100 GHz. The CNS-based nano-oscillators not only offer a feasible pathway toward ultrafast nano-devices, but also hold promise to enable nano-scale energy transduction, harnessing and storage (e.g., from electric to mechanical).

---

[*] Address correspondence to Teng Li, LiT@umd.edu

## 1. Introduction

Significant research progress on graphene in past several years has enabled the exploration of carbon nanoscrolls (CNSs) [1-3], a one-dimensional carbon nanomaterial that is distinct from carbon nanotubes (CNTs). A CNS is formed by rolling up a monolayer graphene into a spiral multilayer nanostructure, whose core size is highly tunable by relative sliding between adjacent layers [4, 5]. In other words, a CNS is topologically open, fundamentally distinct from a tubular CNT, which is topologically closed (e.g., whose core size can only be changed slightly by stretching the carbon-carbon (C-C) bonds). The open and highly tunable structure of CNSs, combining with the exceptional mechanical and electronic properties inherited from the basal graphene [6-9], has inspired an array of novel nano-device applications, such as hydrogen storage medium [10, 11], water and ion channels [12], radially breathing nano-oscillators [13] and translational nano-actuators [14]. In this paper, we demonstrate ultrafast CNS-based axial nano-oscillators that operate at frequencies from 10s GHz to more than 100 GHz, using molecular dynamic (MD) simulations.

Axial nano-oscillators based on multi-walled CNTs (MWCNTs) have been proposed previously [15]. In the proposed MWCNT-based axial nano-oscillator, the ends of the outer tubes of a MWCNT are opened. When the inner tubes are displaced from their original position along the axial direction and then released, the restoring force from the outer tubes pulls the inner ones back. Due to the ultralow friction between the carbon layers, the inner tubes can oscillate along its axial direction, and the natural frequency of the oscillation is estimated to be on the order of GHz [15-17]. Figure 1(a) illustrates a double-walled CNT (DWCNT)-based axial nano-oscillator. Enthusiasm for MWCNT-based axial nano-oscillators aside, the realization of such promising nano-devices hinges upon feasible fabrication techniques. For example, well-controlled opening of the ends of the outer tubes of a MWCNT and chemical treatment of the inner tubes in a MWCNT (e.g., doping or polarization) still remain as significant challenges. As a result, successful fabrication of MWCNT-based axial nano-oscillators has not yet been demonstrated, let alone the exploration of exciting the axial oscillation of such nano-oscillators via external interferences [18-20].

It has been recently demonstrated that a CNT of suitable diameter can initiate the scrolling of a monolayer graphene on a substrate into a CNS [21]. The CNT near the edge of the graphene can help overcome the initial energy barrier for the scrolling of graphene. Once the scrolling is initiated, the graphene can spontaneously roll up into a CNS. The resulting CNS-CNT nanostructure has the two ends of the CNS naturally open and a CNT housed inside the CNS (e.g., Fig. 1(b)). Similar scrolling of a graphene oxide layer initiated by a MWCNT has been experimentally demonstrated recently [22]. As to be detailed later, when the CNT is displaced partially out of the CNS along the axial direction, the van der Waals force acting on the two ends of the CNT is not balanced and the resultant force on the CNT serves as the restoring force to pull the CNT back into the CNS. Given the ultralow CNT-CNS friction similar to the inter-tube friction in a MWCNT, the CNT in the CNS is shown to be able to oscillate at a frequency of 10s GHz. In this paper, we use molecular dynamics simulations to perform systematic investigation of the characteristics of the ultrafast oscillation of the abovementioned CNS-based axial nano-oscillators. We propose a feasible strategy to significantly reduce the energy dissipation of the CNS-based nano-oscillators. We further demonstrate that the CNS-based nano-oscillators can be excited by an external AC electrical field and oscillate at a frequency more than 100 GHz. A distinct advantage of the CNS-based nano-oscillators against the MWCNT-based ones is as follows. The CNT and the basal graphene are fabricated separately before the scrolling process.

For example, the CNT and the graphene can be treated differently and thus possess different features, such as defects, chirality and polarization. These features make it possible to significantly enhance the performance of the CNS-based axial nano-oscillators, as to be detailed later in this paper. With the ever-maturing fabrication technique of high quality graphene, CNS-based axial nano-oscillators hold promise to become a viable approach to achieving nanoscale GHz mechanical oscillators. In particular, the excitation of CNS-based nano-oscillators under external interferences demonstrates their great potential as nanoelectromechanical systems (NEMS) for nanoscale energy transduction (e.g., from electrical and/or magnetic to mechanical), harvesting and storage (e.g., as mechanical oscillation).

## 2. Results and Discussions

### 2.1. CNT-initiated scrolling of graphene into a CNS

The CNS-based axial nano-oscillator depicted in Fig. 1(b) was formed using a 10 nm long (10, 10) single-walled CNT (SWCNT) to initiate the scrolling of a 10 nm by 30 nm graphene along its long (armchair) edge. The formation of the CNS/CNT nanostructure is similar to that described in Ref. [21]. As shown in Fig. 2(a), the graphene is supported by a $SiO_2$ substrate, with a (10, 10) single wall CNT placed along the left edge of the graphene. The substrate is 34 nm long, 14 nm wide and 1 nm thick. In the MD simulations, the C–C bonds in the CNT and CNS are described by the second generation Brenner potential [23], which allows for C-C covalent bond forming and breaking. The non-bonded C-C interaction is described by a Lennard–Jones pair potential [24]. The graphene-substrate interaction is considered in the same way as in Ref. [21]. The MD simulations are carried out using LAMMPS [25] with Canonical Ensemble at 500 K and with time step 1 fs.

Initiated by the CNT, the graphene first separates from the substrate and curls up to wrap the CNT (Fig. 2(b)). Once the overlap between the left edge and the flat portion of the graphene forms (Fig. 2(c)), graphene starts to scroll continuously into a CNS (Fig. 2(d)) with the CNT housed inside. An additional movie file shows the CNT-initiated scrolling of graphene into a CNS (see Additional file 1). Figure 2(e) shows the decrease of the total potential energy due to the graphene wrapping the CNT and further scrolling into a CNS. Our simulations show that there is no appreciable difference between the scrolling of a pristine graphene (without any defects) and that of a graphene with defects. For example, the graphene shown in Fig. 2 has patterned vacancies along three parallel lines, the effect of which is to be detailed later.

## 2.2. Oscillation of a CNT housed inside a naturally formed CNS

The CNS-based nano-oscillator formed by scrolling up a pristine graphene is first equilibrated for 50 ps at 100 K, then the CNT housed inside is assigned a velocity 2.5 Å/ps along its axial direction to initiate the oscillation. In order to constrain the rigid body motion of the nano-oscillator, two rows of carbon atoms along the axial direction on the outermost shell of the CNS are fixed. Figure 3(a) shows the snapshots of the axial oscillation of the CNS/CNT nanostructure at 5 ps, 15 ps, 20 ps, 30 ps, 40 ps and 50 ps, respectively (see Additional file 2 for a video of the oscillation). The simulations are carried out at 100K. The axial motion of the CNT is excluded in the calculation of temperature. Besides the oscillation of the CNT inside the CNS, the CNS itself also oscillates through inter-layer relative sliding in axial direction, initiated by the reaction force from the CNT (i.e., opposite to the restoring force applied on the CNT). The reaction force pulls the inner shells of the CNS to slide outward during the CNT oscillation, thus the CNS itself starts to oscillate accompanying the CNT motion. As a result, the oscillation of the CNS/CNT nanostructure is indeed the coupled CNT oscillation and that of the CNS itself. It needs to be pointed out that the CNS self-oscillation is not only driven by the van der Waals-type reaction force between the CNS and the CNT but also affected by the in-plane shear rigidity of the basal graphene. The different energetic interplays for the CNT oscillation and the CNS self-oscillation lead to a rather irregular coupled oscillation, similar to the axial oscillation observed in a MWCNT [26].

To further decipher the coupled oscillation of the CNS/CNT nanostructure, we define two oscillation amplitudes: the absolute amplitude which is the axial distance from the left end of the CNT to the outermost atom at the left end of the CNS (which is fixed), and the relative amplitude which is the axial distance from the left end of the CNT to the innermost atom at the left end of the CNS (which moves as the CNS oscillates). Figures 3(b) and (d) plot the absolute and relative amplitudes as the function of simulation time, respectively. While the absolute amplitude captures the oscillation of the CNT, the relative amplitude characterizes the coupled oscillation of the CNS/CNT nanostructure, which is more irregular and decays faster. Fast Fourier Transform (FFT) analysis is also performed for the first 500 ps of the oscillation. FFT of the absolute amplitude shows a peak at 29.4 GHz (Fig. 3(c)), which represents the frequency of CNT oscillation. By contrast, FFT of the relative amplitude shows two peaks, 29.4 GHz and 50.9 GHz, respectively (Fig. 3(e)). While the first peak corresponds to the frequency of CNT oscillation inside the CNS, the second peak reveals the frequency of the CNS self-oscillation. The higher frequency of the CNS self-oscillation results from the restoring force contributed by both the non-bonded van der Waals force among carbon layers and the covalent C-C bonding force in the basal graphene.

## 2.3. Oscillation of a CNT housed inside an interlayer-bridged CNS

While the ultrafast oscillation of the CNT inside the CNS at 10s GHz is encouraging, the quick dissipation and rather irregular behavior of the oscillation definitely limit the potential application of CNS-based nano-oscillators as NEMS devices. The quick dissipation and irregular oscillation result from the coupled oscillation, during which the kinetic energy of the CNT is continuously transduced into the self-oscillation of the CNS and then dissipates by the friction due to interlayer sliding. To address this issue, we next demonstrate a feasible and effective strategy to suppress the relative interlayer sliding in the CNS, which can lead to a much more sustainable ultrafast CNS-based nano-oscillator.

Both simulations and experiments have shown that, when a MWCNT is treated by ion irradiation, some carbon atoms can be knocked off, leaving vacancies in the tubes of the MWCNT [27]. Upon heating, the carbon atoms near the vacancies tend to form covalent bonds with other similar carbon atoms in a neighboring tube, driven by the reduction of high-energy dangling bonds of these carbon atoms. As a result, the tubes in the MWCNT are covalently bridged, leading to a significant increase of the inter-tube shear rigidity of the MWCNT. In other words, the relative inter-tube sliding in such a bridged MWCNT involves breaking the covalent C-C bridging bonds, thus is energetically unfavorable. The ion irradiation induced vacancies are also used to facilitate the bridging bond formation among SWCNTs to form CNT bundles [28, 29]. Inspired by these previous studies, next we demonstrate that vacancies can facilitate the formation of interlayer bridging bonds in a CNS, which in turn can effectively suppress the interlayer relative sliding in the CNS.

Instead of using a pristine graphene, we use graphene with patterned vacancies to form a CNS. The vacancies in the graphene are patterned along three parallel lines in the scrolling direction (Fig. 4(a)) to facilitate bridging bond formation after scrolling. In reality, such vacancies can be introduced using focus ion beam to irradiate the graphene along those parallel lines. A SWCNT is used to initiate the scrolling of the aforementioned graphene with patterned vacancies. The carbon atoms at the two ends of the SWCNT are saturated by hydrogen atoms, so that no bridging bonds can be formed between the SWCNT and the CNS. After the scrolling process of the basal graphene with vacancies, the resulting CNS/CNT nanostructure is first heated up from 300 K to 1300 K in 100 ps, then maintained at 1300 K for 1600 ps, and finally cooled down back to 300 K in 100 ps. As shown in Fig. 4(b), interlayer bridging bonds start to form after the temperature reaches 1000 K. The total number of interlayer bridging bonds in the CNS increases as the temperature further increases to and maintains at 1300 K, and gradually saturates (see Additional file 3 for a video of the dynamic process of interlayer bridging bond formation). After cooled down to room temperature, the interlayer bridging bonds formed at high temperature remain in the CNS. Figure 4(c) depicts the end view of the bridged CNS after the heat treatment. Besides the interlayer bridging bonds inside the CNS, bridging bonds are also formed along the unsaturated edges of the CNS (i.e., at the two ends of the CNS and the two edges along its axial direction). No bridging bond is formed between the CNS and the SWCNT with saturated ends.

The oscillation of the SWCNT housed inside the interlayer-bridged CNS is then investigated following the similar procedure used for that of the SWCNT inside the un-bridged CNS. Figure 5(a) shows the snapshots of the axial oscillation of the interlayer-bridged CNS/CNT nanostructure at 25 ps, 35 ps, 45 ps, 55 ps, 65 ps and 75 ps, respectively (see Additional file 4 for a video of the oscillation). No appreciable relative sliding among the CNS layers is found during the oscillation of the CNT. In other words, the self-oscillation of the CNS is effectively suppressed by the interlayer bridging bonds. This is further confirmed by the negligible difference between the absolute amplitude and relative amplitude of the CNT as defined above.

Figures 5(b) plots the absolute amplitude of CNT as a function of simulation time. Compared with the oscillation of the CNT inside an un-bridged CNS, the CNT oscillation inside an interlayer-bridged CNS is much more regular. Also evident in Fig. 5(b) is the slower decay of the oscillation amplitude when compared with Fig. 3(b), which results from the suppression of energy dissipation due to interlayer relative sliding in the CNS. Figure 5(c) plots the peak amplitude of each oscillation cycle and the corresponding oscillation frequency obtained from FFT analysis as a function of simulation time, respectively. The initial frequency of the CNT oscillation is 29.4 GHz when the oscillation amplitude is about 1.15 nm, and the oscillation frequency at 2 ns is 47.0 GHz when the oscillation amplitude is about 0.30 nm. The oscillation frequency increases monotonically as the oscillation amplitude decreases over the time. Such a dependence of oscillation frequency on oscillation amplitude is consistent with the MWCNT-based axial oscillators as reported in earlier studies [30, 31].

We next compare the performance of bridged-CNS-based nano-oscillators with that of MWCNT-based nano-oscillators. Our studies show that, there is negligible difference in the oscillation behaviors between an MWCNT-based nano-oscillator and a DWCNT-based one, if the DWCNT is identical to the two innermost tubes of the MWCNT. Thus, here we report the simulation results of the oscillation behaviors of a (10, 10)/(15, 15) DWCNT, following the similar procedure aforementioned. In order to constrain the rigid body motion of the nano-oscillator, one ring of carbon atoms in the middle of the outer tube of the DWCNT are fixed. The inner tube is assigned a velocity 2.5 Å/ps along its axial direction to initiate the oscillation. The oscillation amplitude, defined as the axial distance from the left end of the inner tube to the left end of the outer tube, is plotted as a function of simulation time in Fig. 6(a). The peak oscillation amplitude of each cycle and the corresponding oscillation frequency as a function of time are shown in Fig.

6(b). While the initial velocity of the inner tube is the same, the resulting initial oscillation amplitude of the DWCNT-based nano-oscillator is slightly smaller than that of the bridged-CNS-based nano-oscillator. Such a difference results from the slight difference in the geometry between the outer tube of the DWCNT (a perfect tube) and the innermost layer of the bridged-CNS (a tube that is cut in axial direction and then slightly displaced radially), leading to a restoring force of the DWCNT-based nano-oscillator modestly larger than that of the bridged-CNS-based one. The difference in the restoring force also explains the relatively higher oscillation frequency of the DWCNT-based nano-oscillator than that of the bridged-CNS-based one for a given oscillation magnitude. Nonetheless, the comparison between Fig. 5(c) and Fig. 6(b) shows that, the bridged-CNS-based nano-oscillator has a modestly slower dissipation rate than the DWCNT-based nano-oscillator. For example, it takes about 1000 ps for the magnitude of DWCNT-based nano-oscillator to decay from 0.9 nm to 0.4 nm, while it takes 1300 ps for the bridged-CNS-based nano-oscillator. We also estimate the quality factor of a nano-oscillator from the evolution of its oscillation amplitude (e.g., Fig. 5b and Fig. 6a) to be

$$Q = \frac{1}{N-10} \sum_{1}^{N-10} \frac{10\pi}{\ln(A_i / A_{i+10})}$$, where $N$ is the total number of oscillation cycles in the MD simulation and $A_i$ denotes the peak amplitude of the $i^{\text{th}}$ cycle. For the bridged-CNS-based nano-oscillator (Fig. 5), $Q \approx 207$ and for the DWCNT-based nano-oscillator (Fig. 6), $Q \approx 192$. Such a comparison of the oscillator performance agrees with the above comparison based on the damping time for a given oscillation amplitude decay. Earlier studies have shown that the translational energy in a DWCNT-based oscillator is mainly dissipated via a wavy deformation in the outer tube undergoing radial vibration [32]. In a bridged CNS, the constraint from the covalent interlayer bridging bonds can largely suppress the radial deformation of all layers in the CNS. In other words, the bridged CNS serves as a thick-walled tubular nanostructure with a

much higher rigidity in both axial and radial directions than a MWCNT. As a result, the axial oscillation of the SWCNT housed inside the bridged CNS is more sustainable than that inside a MWCNT.

## 2.4. Effects of temperature and commensuration on the nano-oscillator performance

To understand the effect of temperature on the performance of the bridged-CNS-based nano-oscillator, Fig. 7 compares the peak oscillation amplitude of each cycle and the corresponding oscillation frequency as a function of time for a bridged-CNS-based nano-oscillator and a DWCNT-based nano-oscillator at 300 K. For both nano-oscillators, the decay of the oscillation magnitude at 300K is modestly faster than that at 100K, while the corresponding oscillation frequency is slightly higher than that at 100K. At higher temperature, the thermal fluctuation of the carbon atoms in the nano-oscillators become more energetic, resulting in rougher surfaces of both the oscillating CNT and the carbon layers of the housing CNT or CNS and therefore increased interlayer friction. Nonetheless, the bridged-CNS-based nano-oscillators still have a modestly slower dissipation rate than the DWCNT-based nano-oscillator at an elevated temperature.

Besides the temperature, the commensuration between the oscillating CNT and the housing CNT or CNS also influences the oscillation performance. It has been shown that the DWCNT-based oscillators with incommensurate inner and outer tubes have lower inter-tube friction force than the commensurate ones, leading to a much slower dissipation rate [30, 33]. To demonstrate the similar effect in bridged-CNS-based nano-oscillators, we replace the (10, 10) SWCNT that is housed inside and commensurate with the interlayer-bridged CNS with an incommensurate (15, 0) SWCNT (whose diameter is very close to (10, 10) SWCNT). Figure 8 reveals that the dissipation rate of the incommensurate bridged-CNS-based nano-oscillator (~0.237 nm/ns) is much slower than that of the commensurate one (~0.429 nm/ns). These results demonstrate an effective strategy to further enhance the performance of bridged-CNS-based nano-oscillators using an incommensurate oscillating SWCNT inside. Our further studies show that the CNT-initiated scrolling of graphene is insensitive to the chirality of the CNT and the basal graphene. This further validates the feasibility of such a strategy since the CNT and the basal graphene can be first synthesized and selected separately and then assembled. By contrast, synthesizing MWCNTs with controlled commensuration among constituent tubes still remains as a grand challenge, let alone leveraging such a strategy to improve the performance of MWCNT-based nano-oscillators.

## 2.5. Oscillation of the CNS/CNT nano-oscillator excited and driven by an external electric field

We further demonstrate that the bridged-CNS-based nano-oscillators can be excited and driven by an external electric field, a crucial feature to enable their potential application in ultrafast NEMS devices. For the MWCNT-based nano-oscillators, it has been proposed that, by inducing net charge [20] or electric dipole [18] into the inner tube, the carbon atoms in the charged/polarized inner tube are subjected to electrostatic capacitive force in an external electric field, which could be potentially used to initialize the oscillation. Controlled charging/polarization of the inner tube of a MWCNT require manipulation with sub-nanometer precision, thus remains rather challenging to achieve experimentally. However, such a strategy can become feasible for bridged-CNS-based nano-oscillators. For example, the SWCNT to be housed inside the interlayer-bridged CNS can be treated to possess net charges or dipoles before used to initiate the scrolling of the basal graphene that remains electrically neutral. Subject to an external AC electric field, the oscillation of the SWCNT housed inside the interlayer-bridged CNS can be initiated and driven by the alternating capacitive force. As a benchmark of such a strategy, Figure 9 shows the oscillation of the bridged-CNS-based nano-oscillator excited and then driven by a square-wave AC electric field with a frequency of 125 GHz. The amplitude of the resulting capacitive force acting on the SWCNT is 0.02 eV/Å per atom. Because such a driving force is much larger than the intensity of the intrinsic van der Waals restoring force between the atoms in the SWCNT and the CNS (~0.0004 eV/Å per atom), the oscillation driven by the external electric field can override the natural oscillation of the bridged-CNS-based nano-oscillator. The slightly asymmetric oscillation amplitude profile in Fig. 9a (e.g., offset by about 0.2 nm) may possibly result from the slightly biased restoring force by the non-uniform atomic structure of the bridged CNS (e.g., due to randomly distributed interlayer bridging bonds). Figure 9(b) shows that the frequency of the resulting oscillation is identical to that of the external AC electric field. Furthermore, there is no appreciable decay in the oscillation amplitude, whose peak value in each oscillation cycle only fluctuates within 5%. In other words, the oscillation driven by the external electric field is highly sustainable. Our further studies show that the resulting oscillation of the bridged-CNS-based nano-oscillator can be further fine tuned in a certain range under an external AC electric field of suitable frequency and magnitude. These explorations further demonstrate the potential to leverage CNS-based nano-oscillators to convert the electric energy of an external AC field into mechanical energy in the form of ultrafast oscillation. With proper treatment of the oscillating CNT, the above strategy can be potentially adapted to transduce and harvest electromagnetic and thermal energy into ultrafast mechanical oscillation [16, 34].

## 3. Conclusions

To conclude, we demonstrate a new type of ultrafast axial nano-oscillators based on CNS. Such a nano-oscillator consists of a SWCNT that is housed inside a CNS, which can be feasibly formed by the SWCNT-initiated scrolling of a basal monolayer graphene. The unique topological structure of the CNS-based nano-oscillator offers a viable pathway to fabricating ultrafast axial nano-oscillators, addressing a significant challenge that still remains for the previously proposed MWCNT-based axial nano-oscillator. We propose an effective and feasible strategy to reduce the oscillation dissipation of the CNS-based nano-oscillators by introducing interlayer bridging bonds in the CNS. The performance of the resulting bridged-CNS-based nano-oscillators is comparable or modestly better than the MWCNT-based ones. We further demonstrate the highly sustainable oscillation of the bridged-CNS-based nano-oscillators that can be excited and driven by an external AC electric field. With the ever maturing fabrication of high quality monolayer graphene and nanofabrication technique of patterning nanoscale building blocks, we envision a novel approach to harnessing and storing energy at nanoscale and over large area, enabled by distributing CNS-based nano-oscillators on an electronic surface.

**Abbreviations**

CNT: carbon nanotube; CNS: carbon nanoscroll, SWCNT: single-walled carbon nanotube; DWCNT: double-walled carbon nanotube; MWCNT: multi-walled carbon nanotube; MD: molecular dyanimics; GHz: gigaherz.

**Acknowledgements**

This work is supported by National Science Foundation (grant #1069076), University of Maryland General Research Board Summer Research Award, and Maryland NanoCenter at the University of Maryland, College Park. ZZ also thanks the support of A. J. Clark Fellowship and

**Authors' contributions**

TL designed and supervised research; ZZ carried out simulations; TL and ZZ analyzed data; and

TL and ZZ wrote the paper.

**Competing interests**

The authors declare that they have no competing interests.

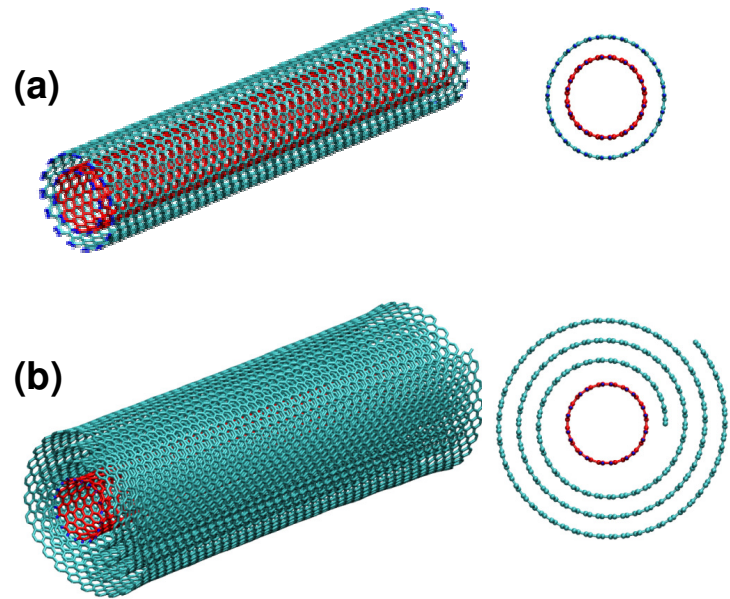

**Figure 1.** Perspective view (left) and end view (right) of (a) a DWCNT and (b) a CNS with a SWCNT housed inside. When displaced from its equilibrium position along axial direction, the inner tube (red) can oscillate inside the outer tube (cyan) or CNS (cyan), at GHz frequency.

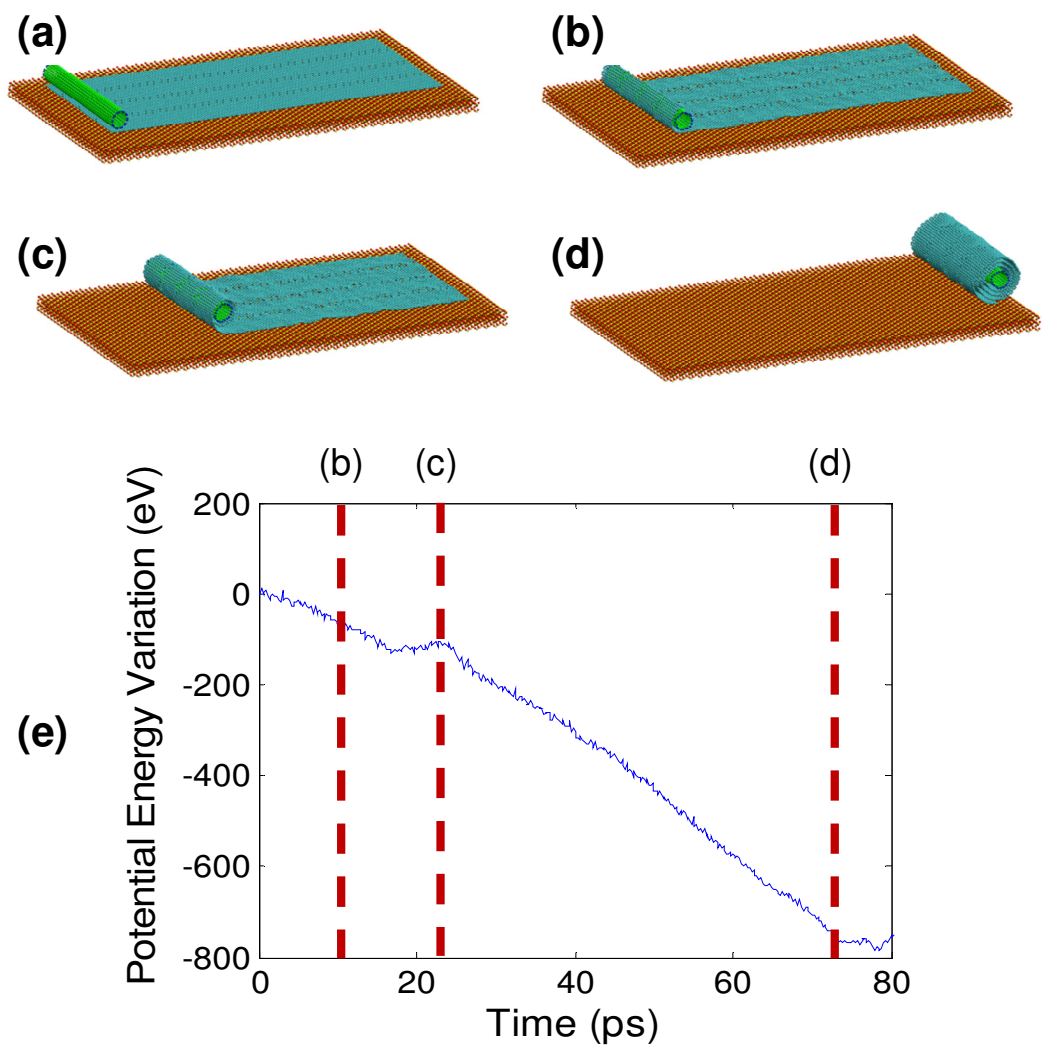

**Figure 2**. (a - d) Snapshots of the graphene scrolling into a CNS, initiated by a (10, 10) SWCNT, before equilibration, at 10 ps, 22 ps, and 76 ps, respectively. (e) The variation in the total potential energy of the system as a function of simulation time.

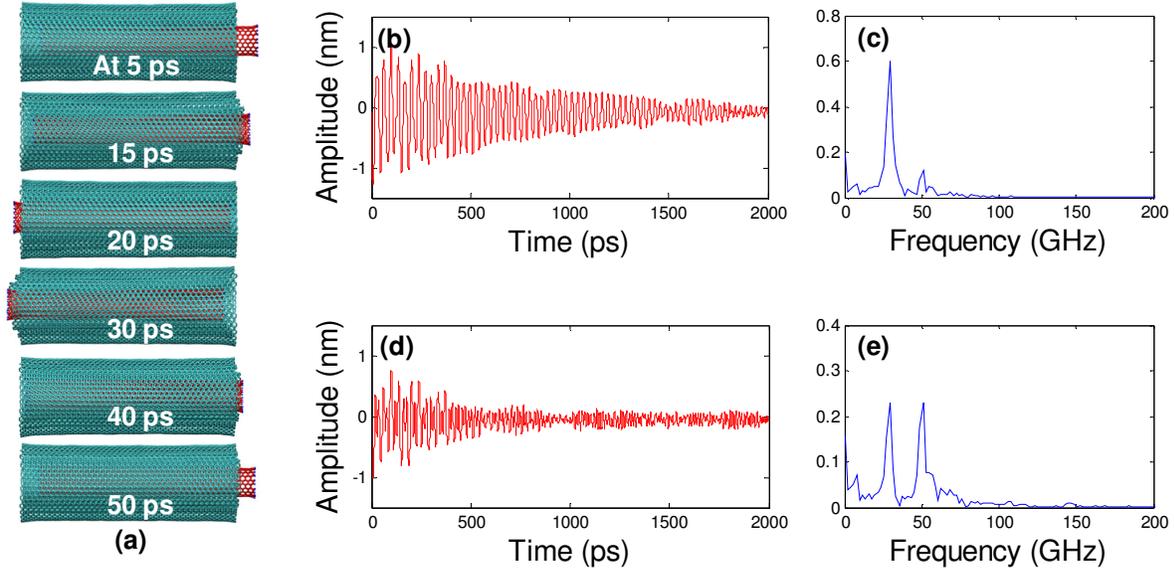

**Figure 3.** (a) Snapshots of the axial oscillation of the CNS-based nano-oscillator at 5 ps, 15 ps, 20 ps, 30 ps, 40 ps and 50 ps, respectively. Note the coupled axial oscillations of the CNT and the CNS itself. The evolution of (b) the absolute amplitude and (d) the relative amplitude of the CNT oscillation inside the CNS, as a function of simulation time, respectively. The Fast Fourier Transform (FFT) analysis of the absolute amplitude (c) and the relative amplitude (e) for the first 500 ps reveals a frequency of the oscillation of the CNT inside the CNS (29.4 GHz) and that of the oscillation of the CNS itself (50.9 GHz), respectively. The simulations are carried out at 100K.

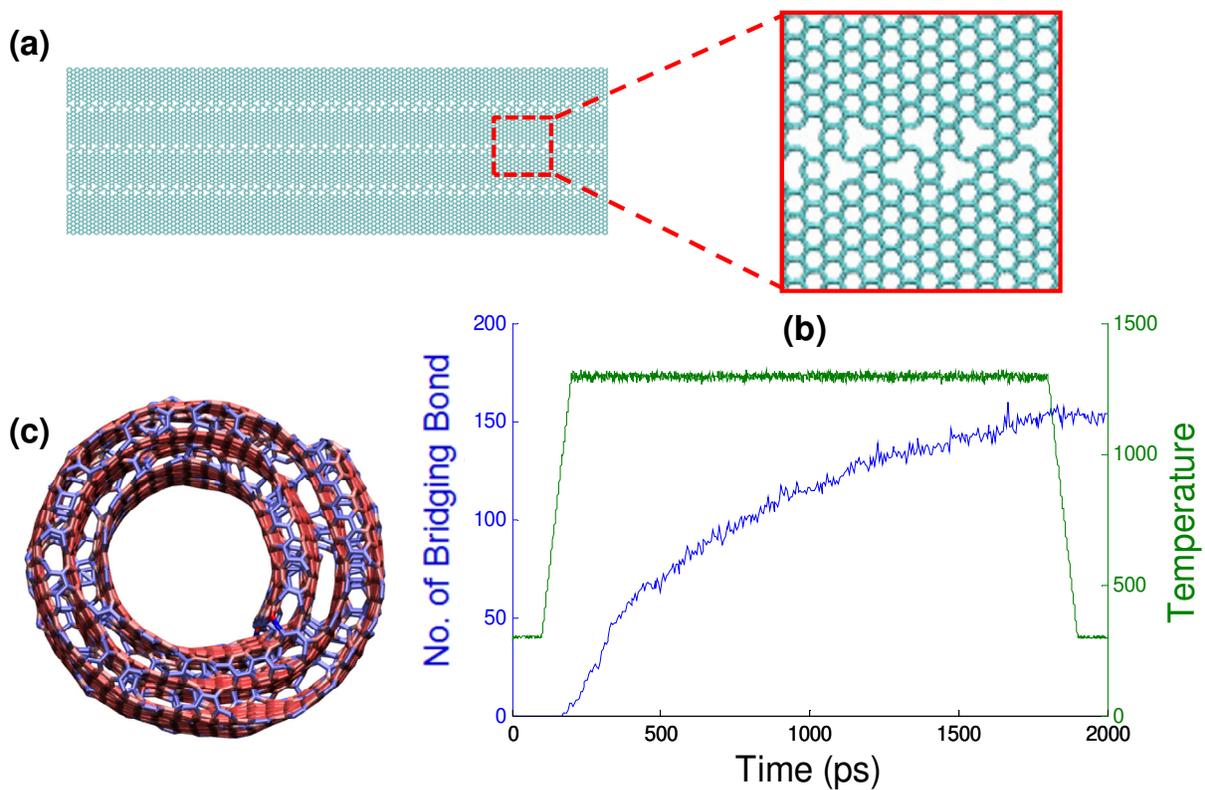

**Figure 4.** (a) The graphene with patterned vacancies. (b) The evolution of the number of interlayer bridging bond in the CNS and the temperature change as a function of time, respectively. Note that the bridging bonds remain after cooling down to room temperature. (c) The end view of the interlayer-bridged CNS after the heat treatment. The color shades represent potential energy level of the carbon atoms. Here the SWCNT housed inside the CNS is not shown for visual clarity.

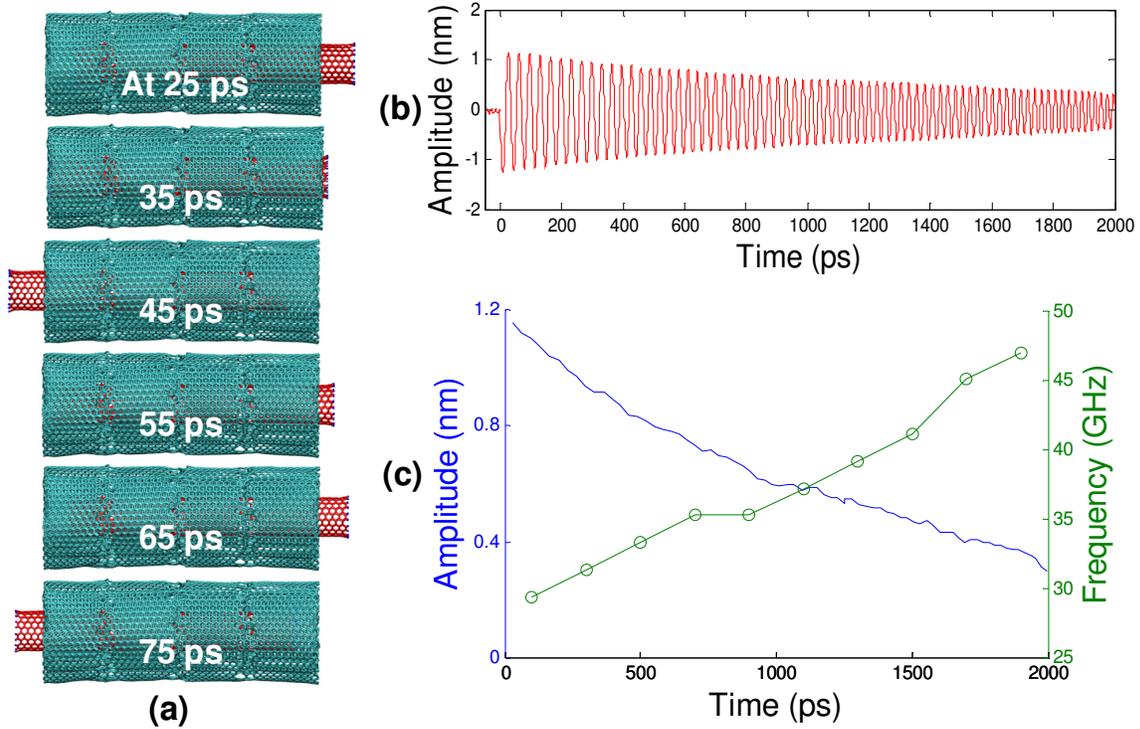

**Figure 5.** (a) Snapshots of the axial oscillation of the bridged-CNS-based nano-oscillator at 25 ps, 35 ps, 45 ps, 55 ps, 65 ps and 75 ps, respectively. Note the oscillation of the CNS itself is fully constrained by the interlayer bridging bonds. (b) The evolution of CNT oscillation amplitude. (c) The peak amplitude of each oscillation cycle and the corresponding oscillation frequency as a function of time, respectively. The simulations are carried out at 100K.

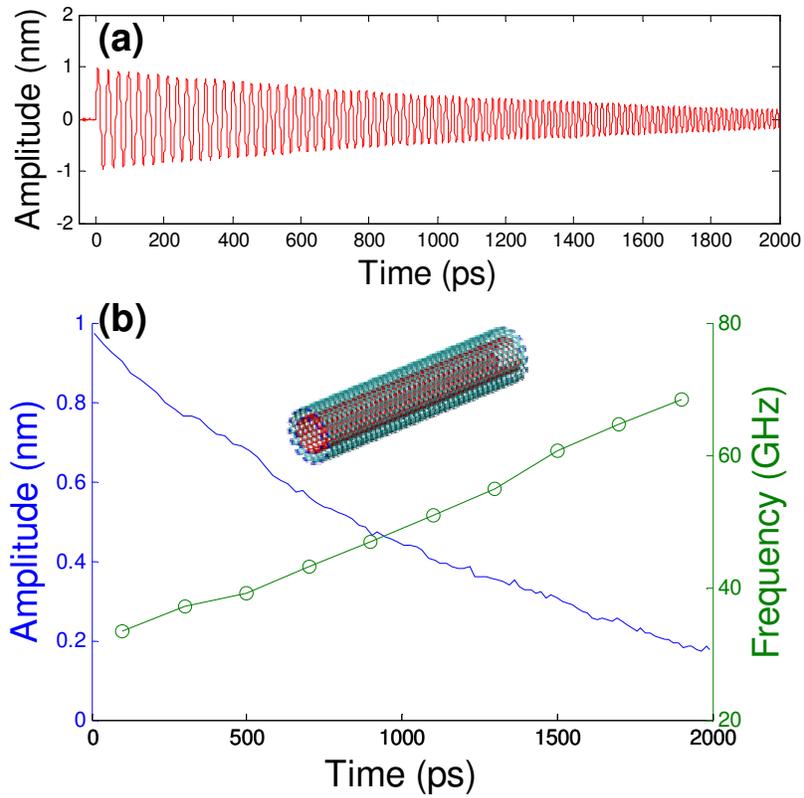

**Figure 6.** (a) The evolution of the oscillation amplitude of the inner tube of a (10, 10)/(15, 15) DWCNT. (b) The peak oscillation amplitude of each cycle and the corresponding oscillation frequency as a function of time, respectively. The simulations are carried out at 100K.

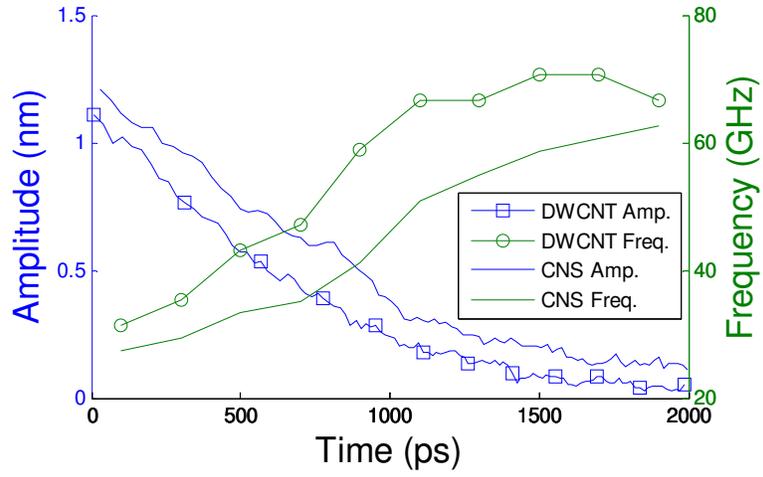

**Figure 7.** The comparison between the bridged-CNS-based nano-oscillator and the DWCNT-based nano-oscillator at 300K.

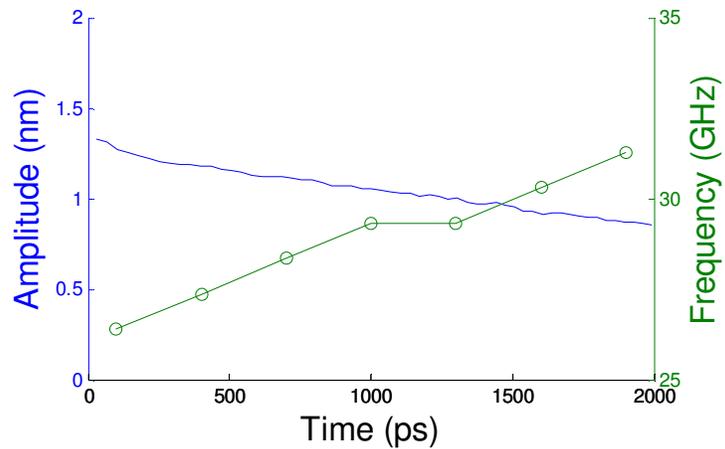

**Figure 8.** The peak oscillation amplitude of each cycle and the corresponding oscillation frequency as a function of time for a (15, 0) SWCNT inside the interlayer-bridged CNS, respectively. The simulations are carried out at 100K.

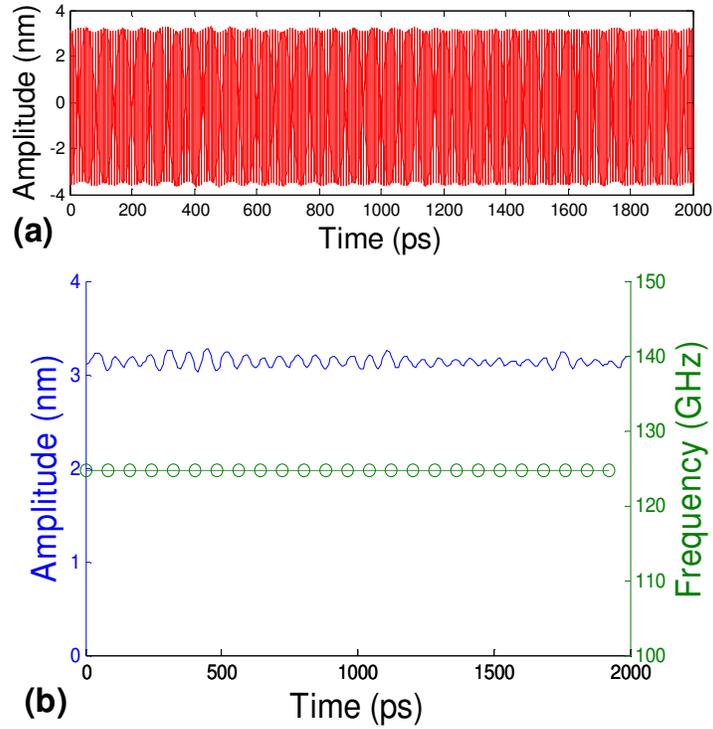

**Figure 9.** (a) The oscillation of the CNT in the interlayer-bridged CNS excited and driven by an external electrical field with an ac frequency of 125 GHz. (b) The peak oscillation amplitude of each cycle and the corresponding oscillation frequency as a function of time. The external ac electrical field can override the natural frequency of the CNS-based nano-oscillator. There is no appreciable decay of peak oscillation amplitude. The simulations are carried out at 100 K.